\documentstyle[epsf,aps,preprint,prb]{revtex}
\draft
\begin{document} 

\title{\bf First-Principles Study of Carbon Monoxide Adsorption on Zirconia-Supported Copper} 
\author{Eric J. Walter}
\address{Department of Chemistry and Laboratory for Research on the Structure of Matter, \\ 
University of Pennsylvania, Philadelphia, PA 19104-6323.}  
\author{Steven P. Lewis}
\address{Department of Physics and Astronomy,
University of Georgia, Athens, GA 30602-2451.}
\author{Andrew M. Rappe}
\address{Department of Chemistry and Laboratory for Research on the Structure of Matter, \\ 
University of Pennsylvania, Philadelphia, PA 19104-6323.}  

\date{\today}
\maketitle

\begin{abstract}

We have calculated the adsorption energy of carbon monoxide on a
monolayer of copper adsorbed on the (111) face of cubic zirconia.  We
investigate the structural parameters of three phases of bulk zirconia
(cubic, tetragonal, and monoclinic) and find excellent agreement with
experiment.  We have also analyzed the structural relaxation of both
the stoichiometric and reduced (111) surfaces of cubic zirconia
($c$-ZrO$_2$). For adsorption of copper on $c$-ZrO$_2$, we find that
the preferred binding site is atop the terminal oxygen atom, favored
by 0.3 eV over other high symmetry sites.  We compare CO adsorption on
zirconia-supported copper to the results of carbon monoxide on
copper (100) (S. P. Lewis and A. M. Rappe, J. Chem. Phys. {\bf 110},
4619, (1999).) and show that adsorption on oxide-supported copper is
over 0.2 eV more stable than adsorption on the bare surface.

\noindent
\end{abstract}


\newpage
\section{INTRODUCTION}
\label{sect:intro}

Oxide-supported transition metals have received much attention due to
their utility in automotive catalysis and other catalytic processes.
Use of an oxide as a support material for transition metals has clear
economic benefits because this reduces the amount of costly noble
metals (e.g., Rh, Pt) that need to be used in functioning catalysts.
Moreover, the ability of the support material to contribute oxygen to
chemical reactions can significantly enhance many catalytic
mechanisms.  Zirconium dioxide (or zirconia) is a technologically
important catalytic support medium.  This material has many other
applications, as well, including gas sensors, solid fuel cells, and
high durability coatings.

Copper has recently been identified as an important catalytic agent
for NO reduction and CO oxidation.\cite{SR,DN,DB,CD,B,YO1,YO2} For
instance, Cu-ion exchanged zeolites such as
Cu/ZSM-5\cite{SR,DN,DB,CD}, alumina-supported CuO\cite{B} and
ZrO$_2$-supported copper\cite{YO1,YO2} all exhibit high activity for
the catalytic promotion of NO$_x$ reduction by CO (forming N$_2$ and
CO$_2$).  An attractive quality of the Cu/ZrO$_2$ system is its
unusually high catalytic activity for the NO-CO reaction at very low
temperatures (100-200$^{\circ}$C).\cite{YO1,YO2}

Because of the interest in Cu/ZrO$_2$ as a part of the next generation
of automotive catalysts, we have performed a series of {\it ab initio}
calculations to study the energetics of molecular adsorption on this
surface.  Although there have been previous theoretical studies of
bulk
ZrO$_2$\cite{Jansen,Orlando,Rosenbauer,Stef,Wilson,Kralik,Dewhurst,Kahn}
and a recent thorough investigation of ZrO$_2$ surfaces\cite{Carter},
this investigation is the first to study molecular adsorption onto
metal films on this surface.  To the best of our knowledge, this is
also the first {\it ab initio} study of oxide-supported metal
chemisorption.  A monolayer of copper was chosen for several reasons.
The monolayer geometry can be conveniently determined theoretically,
and we believe that it can be can be reproducibly prepared
experimentally.  Adsorption onto a monolayer of copper also forms the
starting point for systematic studies of clusters of atoms on
surfaces.  Furthermore, interactions of molecules with individual
atoms in a copper monolayer makes contact with zeolite systems.  To
aid in interpretation, we also make a direct comparison between
adsorption onto Cu/ZrO$_2$ and onto the bare Cu(100) surface.\cite{LR}

\section{METHOD}

All calculations in this study use density functional
theory\cite{HK,KS} (DFT) within the plane-wave pseudopotential\cite{Payne}
method to determine all structural parameters and energies.  Optimized
pseudopotentials\cite{Rappe} for all elements were constructed to be 
well converged for a 50 Ry plane wave cut-off energy.
 
For structural parameters, the local density approximation
(LDA)\cite{PZ} of DFT gives highly accurate results for most bulk and
surface systems. Typically, the optimal computed parameters are within
1\% of the experimental values.  However, for adsorption energies, the
generalized gradient approximation (GGA)\cite{PBE} typical yields
significantly more accurate results.  To obtain GGA results for the
adsorption energies of copper and carbon monoxide on the substrate we
take a hybrid approach.  First, the structure is fully relaxed within
the LDA (i.e., the atoms are moved until all atomic forces are less
than 0.01 eV/\AA).  Second, the GGA energy of the optimal
configuration is obtained from the ground-state charge density of the
LDA calculation.  This hybrid approach yields significantly more
accurate adsorption energies than LDA calculations alone; for most
systems, this approach gives results very close to fully
self-consistent GGA calculations.\cite{Sch}

For bulk systems, Monkhorst-Pack\cite{MP} special $k$-point sets were
used for Brillouin zone integrations.  For the bulk cubic system,
calculations using 10 irreducible $k$-points gave convergence error of
less than 1 meV per unit cell.  Similar convergence criteria were used
for the other bulk systems.  For calculations involving the (111)
surface unit cell, we used the Ram\'{i}rez-B\"{o}hm\cite{RB} $k$-point
sampling method and found that convergence was obtained with 3
irreducible $k$-points.

\section{RESULTS and DISCUSSION}

\subsection{Bulk Zirconia}

The ground-state phase of zirconia---{\it baddeleyite}---has a very
complex monoclinic structure ($m$-ZrO$_2$) with nine internal degrees
of freedom and four formula units per primitive cell.  Each Zr cation is
7-fold coordinated by oxygen in the monoclinic phase.  The structure
can be viewed as alternating layers of Zr and O with the coordination
of the O atoms alternating between 3 and 4 from one oxygen layer to
the next.

At about 1400 K, a first-order martensitic phase transition occurs,
yielding the tetragonal phase ($t$-ZrO$_2$).  The tetragonal form of
zirconia can be viewed as a simple distortion of the cubic fluorite
structure, with alternating columns of oxygen atoms along one
crystallographic axis shifting upward or downward by an amount $d_z$.
This structure is described by two lattice constants, $a$ and $c$, and
it has two formula units per unit cell.

Above about 2650 K, zirconia assumes the cubic fluorite structure
($c$-ZrO$_2$).  In this phase there are only two degrees of freedom:
the lattice constant, $a$, and an internal coordinate, $u$, reflecting
the positions of the oxygen atoms along the body diagonal of the cubic
cell.  For the ideal fluorite structure, the value of this coordinate
is 0.25.  The cubic structure can be stabilized at room temperature by
incorporation of a few percent of Y$_2$O$_3$.  Addition of cation
impurities also improves the thermochemical properties substantially,
giving the cubic phase extremely high strength and thermal-shock
resistance.

Table 1 shows the optimized structural parameters for the monoclinic,
tetragonal and the cubic phase of bulk zirconia (see Fig. 1).  All
computed values are in excellent agreement with experiment.\cite{Stef2}

\subsection{Cubic Zirconia (111) Surface}

Table 2 shows the relaxation data for two (111) surfaces of {\it
c}-ZrO$_2$: the stoichiometric and the fully reduced (i.e., top layer
of oxygen removed) surfaces.  The stoichiometric surface shows an
outward relaxation of the top oxygen-zirconium spacing and a
significant reduction in the oxygen-oxygen interlayer spacing between
the two outermost layers of ZrO$_2$ formula units.  For the reduced
surface, removal of the top layer of oxygen changes these relaxations
dramatically.  We predict a huge (24.4\%) inward relaxation of the top
zirconium-oxygen spacing and a reversal in the relaxation of the
oxygen-oxygen spacing between the top two layers of formula units.
Our calculations show a large energetic cost of 9.4 eV/atom for
completely reducing the (111) surface.  The possibility of surface
reconstructions has not been considered in this study.  A detailed
study of many zirconia surfaces can be found in reference 16.

\subsection{Copper on {\it c}-Zirconia}

Before investigating the chemisorption of CO on zirconia supported
copper, we first determine the preferred binding site for copper on
(111) {\it c}-ZrO$_{2}$.  There are three high-symmetry positions on
this surface: the top oxygen site (O1), the three-fold hollow site
directly above the zirconium atom (Zr), and the hollow site above the
subsurface oxygen atom in the top ZrO$_{2}$ formula unit (O2).  Table
3 shows the GGA binding energies for a monolayer of copper centered on
each of these three sites.  As shown in the table, the O1 site is
preferred by about 0.3 eV/atom over the other possible sites.

It is important to compare the binding energy of copper on the {\it
c}-ZrO$_2$ surface with the cohesive energy of bulk copper.  Since the
cohesive energy is nearly twice the adsorption energy of copper on
{\it c}-ZrO$_2$, the oxide-supported copper monolayer is, at best, a
metastable configuration.  Annealing this monolayer would result in
the formation of copper particles on the oxide surface.

\subsection{CO on Cu on ZrO$_2$(111) with comparison to CO on Cu(100) }

With the preferred binding site of copper determined, the binding
energy of carbon monoxide on the oxide-supported monolayer of copper
can now be calculated.  Table 4 shows the GGA binding energy and the
relaxed structural parameters of carbon monoxide on the oxide
supported copper monolayer.  It is illustrative to compare the binding
energy of carbon monoxide on zirconia-supported copper to the binding
energy of a half monolayer of carbon monoxide on the Cu(100)
surface\cite{LR}.  As seen in Table 4, the nearest-neighbor distance
of CO molecules is very similar for these systems.

Our computations reveal that the binding energy of CO on
zirconia-supported copper is nearly 0.2 eV greater than that of CO on
the Cu (100) surface.  This is most likely due to charge transfer from
the copper atoms to the zirconia substrate.  It has been shown
experimentally\cite{BD} that the Cu-CO chemisorption bond is dative,
resulting from charge transfer from the weakly anti-bonding 5$\sigma$
orbital of the CO molecule to the metal conduction bands.  This
process is therefore enhanced due to the presence of the strongly
electronegative oxide substrate.

\section{Conclusions}

We have computed the bulk parameters of monoclinic, tetragonal and
cubic zirconia which are in excellent agreement with experiment.  Our
surface relaxation data for the stoichiometric (111) {\it c}-ZrO$_2$
surface shows an outward relaxation of the top layer of oxygen.  This
may be related to the propensity of the ZrO$_2$ top-site oxygen atoms
to participate in surface chemical reactions.  Furthermore, we have
found that the preferred binding site for copper on the $c$-ZrO$_2$
(111) surface, by 0.3 eV/atom, is atop the surface oxygen atom.  We
have calculated the adsorption energy of CO on an oxide-supported
monolayer of copper.  Comparing our results to that of CO on Cu (100),
we find that the presence of the support increases the binding energy
by over 0.2 eV/molecule.

\acknowledgments

This work was supported by the Laboratory for Research on the
Structure of Matter as well as NSF grant DMR 97-02514.  The authors
would also like to thank the Alfred P. Sloan Foundation for support.
Computational support was provided by the National Center for
Supercomputing Applications and the San Diego Supercomputer Center.

\newpage
\begin{figure}[p]
\epsfysize=2.5in
\centerline{\epsfbox[0 0 330 435]{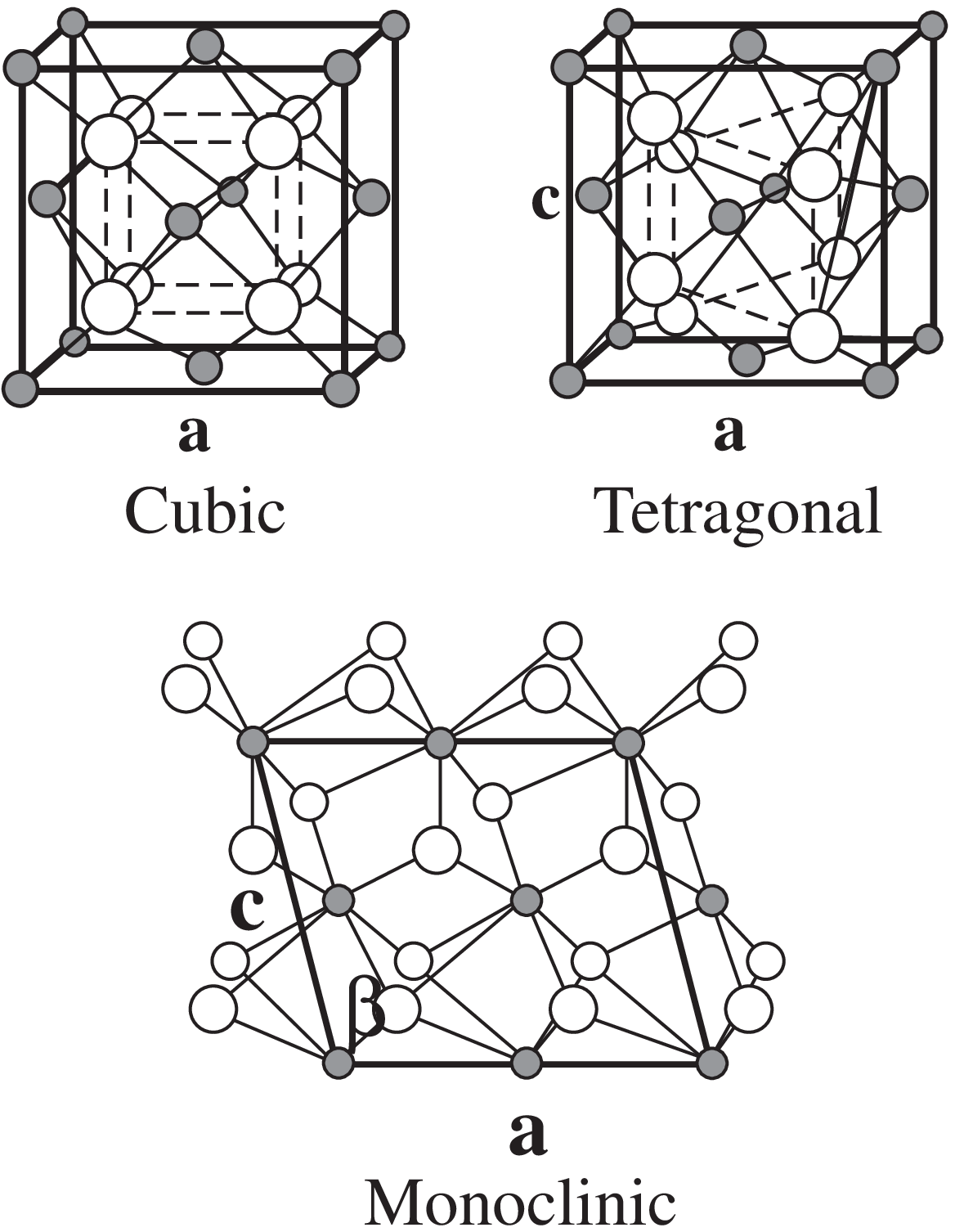}}
\caption{Unit cells for cubic, tetragonal monoclinic ZrO$_2$.  Oxygen
atoms are white, zirconium atoms are grey.}
\end{figure}

\begin{figure}[p]
\epsfysize=1.5in
\centerline{\epsfbox[0 0 530 220]{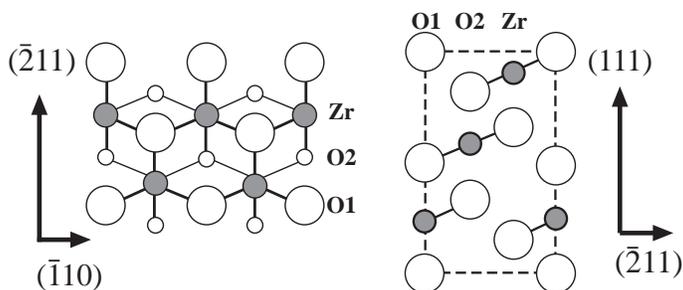}}
\caption{Top and side view of the (111) $c$-ZrO$_2$ surface.  Copper
atoms were adsorbed to the sites listed (O1, O2 and Zr).}  
\end{figure}

\begin{table}[p]
\begin{center}
\caption{Optimized structural parameters for cubic, tetragonal
and monoclinic phases of zirconia. }
\vspace{5mm}
\begin{tabular}{lcc}
Parameter & This Work & Experiment\tablenote[1]{Reference 27}\\ 
\multicolumn{3}{c}{\bf Cubic}\\
\hline
$a$ & 5.043 & 5.090\\
$u$ & 0.260 & 0.275\\
\multicolumn{3}{c}{\bf Tetragonal}\\
\hline
$a$ & 5.047 & 5.050\\
$c$ & 5.127 & 5.182\\
$d_z$ & 0.043 & 0.0574\\
\multicolumn{3}{c}{\bf Monoclinic}\\
\hline
$a$ & 5.226 & 5.317\\
$b$ & 5.082 & 5.150\\
$c$ & 5.229 & 5.212\\
$\theta$ & $99.18^{\circ}$ & $99.23^{\circ}$\\
$x_{\rm Zr}$ & 0.2107 & 0.2083\\
$y_{\rm Zr}$ & 0.2778 & 0.2754\\
$z_{\rm Zr}$ & 0.0417 & 0.0395\\
$x_{\rm O1}$ & 0.3303 & 0.3317\\
$y_{\rm O1}$ & 0.0771 & 0.0700\\
$z_{\rm O1}$ & 0.3500 & 0.3447\\
$x_{\rm O2}$ & 0.4824 & 0.4792\\
$y_{\rm O2}$ & 0.4477 & 0.4496\\
$z_{\rm O2}$ & 0.7588 & 0.7569

\label{tab:bulk_data}
\end{tabular}
\end{center}
\end{table}

\begin{table}[p]
\begin{center}
\caption{Relaxation data for the (111) surface of cubic zirconia, stoichiometric and reduced.  Data are listed as the percent
deviation from the bulk terminated interlayer spacing.}
\vspace{5mm}
\begin{tabular}{ldd}
{\bf Spacing}& {\bf Stoichiometric}& {\bf Reduced}\\
\hline
$\Delta_{\rm O-Zr}$ & +5.89  & ---  \\
$\Delta_{\rm Zr-O}$ & -0.03  &-24.4 \\
$\Delta_{\rm O-O}$ & -3.88  &+4.70 \\
$\Delta_{\rm O-Zr}$ & +1.14  &+0.70 \\
$\Delta_{\rm Zr-O}$ & +0.62  &-1.63 \\
$\Delta_{\rm O-O}$ & -0.12  &+0.08 
\label{tab:relax}
\end{tabular}
\end{center}
\end{table}

\begin{table}[p]
\begin{center}
\caption{Binding energies of copper at various sites on the (111) surface of cubic zirconia}
\vspace{5mm}
\begin{tabular}{ld}
{\bf Site}& {\bf Binding Energy(eV/atom)} \\
\hline
O1 & 1.8  \\
Zr & 1.5 \\
O2 & 1.3 \\
Bulk Cu & 3.5 
\label{tab:Cu data}
\end{tabular}
\end{center}
\end{table}

\begin{table}[p]
\begin{center}
\caption{Structural data and binding energies of CO on Cu (100) and zirconia supported copper}
\vspace{5mm}
\begin{tabular}{ldd}
{\bf Parameter}& {\bf CO-Cu-ZrO$_2$}& {\bf CO-Cu(100)} \\
\hline
Binding energy(eV) & 0.86 & 0.65\\
CO-CO distance (\AA)  & 3.61 & 3.56\\
C-O bond length   (\AA)  & 1.140 & 1.139\\ 
Cu-CO bond length (\AA)  & 1.801 & 1.852
\label{tab:CO data}
\end{tabular}
\end{center}
\end{table}


\begin{thebibliography}{MM}

\bibitem{SR} S. Recchia, C. Dossi, A. Fusi, R. Psaro, R. Ugo and
G. Moretti, Chem. Commun. {\bf 19}, 1909, (1997).

\bibitem{DN} D. Nachtigallova, P. Nachtigall, M. Sierka and J. Sauer,
Phys. Chem. Chem. Phys. {\bf 1}, 2019, (1999).

\bibitem{DB} D. Biglino, H. Li, R. Erickson, A. Lund, H. Yahiro and
M. Shiotani, Phys. Chem. Chem. Phys. {\bf 1}, 2887, (1999).

\bibitem{CD} C. Dossi, S. Recchia, A. Pozzi, A. Fusi, V. Dalsanto and
G. Moretti, Phys. Chem. Chem. Phys. {\bf 1}, 4515, (1999).

\bibitem{B} G. L. Bauerle, G. R. Service and K. Nobe,
Ind. Eng. Chem. Prod. Res. Dev. {\bf 11}, 54, (1972).

\bibitem{YO1} Y. Okamato, T. Kubota, H. Gotoh, Y. Ohto, H. Aritani,
T. Tanaka and S. Yoshida, J. Chem. Soc., Faraday Trans. {\bf 94},
3743, (1998).

\bibitem{YO2} Y. Okamato, H. Gotoh, Y. Ohto, H. Aritani, T. Tanaka and
S. Yoshida, J. Chem. Soc., Faraday Trans. {\bf 93}, 3879, (1997).

\bibitem{Jansen} H. J. F. Jansen, Phys. Rev. B {\bf 43}, 7267, (1991).

\bibitem{Orlando} R. Orlando, C. Pisani, C. Roetti, and
E. Stefanovich, Phys. Rev. B {\bf 45}, 592, (1992).

\bibitem{Rosenbauer} M. Rosenbauer and H. J. F. Jansen, Phys. Rev. B
{\bf 47}, 16148, (1993).

\bibitem{Stef} E. V. Stefanovich, A. L. Shluger, and C. R. A. Catlow,
Phys. Rev. B {\bf 49}, 11560, (1994).

\bibitem{Wilson} M. Wilson, U. Sch\"{o}nberger, and M. W. Finnis,
Phys. Rev. B {\bf 54}, 9147, (1996).

\bibitem{Kralik} B. Kr\'{a}lik, E. K. Chang, and S. G. Louie,
Phys. Rev. B {\bf 57}, 7027, (1998).

\bibitem{Dewhurst} J. K. Dewhurst and J. E. Lowther, Phys. Rev. B {\bf
57}, 741, (1998).

\bibitem{Kahn} M. S. Kahn, M. S. Islam, and D. R. Bates,
J. Mater. Chem {\bf 8}, 2299, (1998).

\bibitem{Carter} A. Christensen and E. A. Carter, Phys. Rev. B {\bf
58}, 8050, (1998).

\bibitem{LR} S. P. Lewis and A. M. Rappe, J. Chem. Phys {\bf 110},
4619, (1999).

\bibitem{HK} P. Hohenberg and W. Kohn, Phys. Rev. {\bf 136} (1964)
B864.

\bibitem{KS} W. Kohn and L. J. Sham, Phys. Rev. {\bf 140} (1965)
A1133.

\bibitem{Payne} M. C. Payne, M. P. Teter, D. C. Allan, T. A. Arias,
and J. D. Joannopoulos, Rev. Mod. Phys. {\bf 64} (1992) 1045.

\bibitem{Rappe} A. M. Rappe, K. M. Rabe, E. Kaxiras, and
J. D. Joannopoulos, Phys. Rev. B {\bf 41} (1990) 1227.

\bibitem{PZ} J. Perdew and A. Zunger, Phys. Rev. B {\bf 23} (1981)
5048.

\bibitem{PBE} J. P. Perdew, K. Burke and M. Ernzerhof,
Phys. Rev. Lett. {\bf 77} (1996) 18.

\bibitem{Sch} M. Fuhcs, M. Bockstedte, E. Pehlke and M. Scheffler,
Phys. Rev. B {\bf 57}, 2134, (1998)

\bibitem{MP} H. J. Monkhorst and J. D. Pack, Phys. Rev. B {\bf 16},
5188 (1976).

\bibitem{RB} R. Ram\'{i}rez and M. C. B{\"{o}}hm, Int. J. of
Quant. Chem. {\bf 30}, 391, (1986).

\bibitem{Stef2} Experimental data taken from C. J. Howard, R. J. Hill,
and B. E. Riechert, Acta Crystallogr. Sect. B {\bf 44} 116 (1988) and
P. Aldebert and J. -P. Traverse, J. Am. Ceram. Soc. {\bf 68}, 34
(1985) as presented in table V of E. V. Stefanovich, A. L. Shluger,
and C. R. A. Catlow, Phys. Rev. B {\bf 49}, 11560, (1994). 

\bibitem{BD} E. Borguet and H.-L. Dai, J. Chem. Phys. {\bf 101}, 9080,
(1994).

\end{thebibliography}
\end{document}